\documentclass[manuscript]{aastex62}

\usepackage[T1]{fontenc}

\usepackage{bm}        
\usepackage{amssymb, amsmath}   
\usepackage{cancel}
\usepackage{color}

\newcommand\aastex{AAS\TeX}

\DeclareMathAlphabet{\mathsfit}{\encodingdefault}{\sfdefault}{m}{sl}
\SetMathAlphabet{\mathsfit}{bold}{\encodingdefault}{\sfdefault}{bx}{sl}

\received{****}
\revised{****}
\accepted{****}

\shorttitle{\aastex\ Heavy ion acceleration}
\shortauthors{Fu et al.}

\begin{document}

\title{Heating of Heavy Ions in Low-beta Compressible Turbulence}%

\correspondingauthor{Xiangrong Fu}
\email{sfu@newmexicoconsortirum.org}

\author{Xiangrong Fu}
\affil{New Mexico Consortium, Los Alamos, NM 87544}
\author{Fan Guo}
\affil{Los Alamos National Laboratory, Los Alamos, NM 87545}
\affil{New Mexico Consortium, Los Alamos, NM 87544}
\author{Hui Li}
\affil{Los Alamos National Laboratory, Los Alamos, NM 87545}
\author{Xiaocan Li}
\affil{Los Alamos National Laboratory, Los Alamos, NM 87545}

\begin{abstract}
Enhancement of minor ions such as $^3$He and heavy ions in flare-associated 
solar energetic particle (SEP) events remains one of the major puzzles in heliophysics. In this work, we use 3D hybrid simulations (kinetic protons and fluid electrons) to investigate particle energization in a turbulent low-beta environment similar to solar flares. It is shown that in this regime the injected large-amplitude Alfv\'en waves develop into compressible and anisotropic turbulence, which efficiently heats thermal ions of different species. We find that temperature increase of heavy ions is inversely proportional to the charge to mass ratio, which is consistent with observations of impulsive SEP events. Further analysis reveals that ions are energized by interacting with nearly perpendicular magnetosonic waves near proton inertial scale. 
\end{abstract}



\section{\label{sec:intro}Introduction}

Solar energetic particle (SEP) events are abrupt enhancement of high-energy (10 keV - GeV) charged particle fluxes by several orders of magnitude during solar activities \citep{reames_sp_2017}. While the majority of SEPs are protons and electrons, minor ions such as helium, oxygen, carbon, and iron have also been measured routinely. In fact, information on the relative abundance of different elements and their isotopes provides insights into the mechanisms that produce SEPs.  

Solar energetic particle events are often organized into two categories: gradual and impulsive events, based on the duration of enhanced fluxes. The gradual SEP events are usually correlated with interplanetary shocks driven by coronal mass ejections, while the impulsive SEP events are thought to be driven mainly by magnetic reconnection processes, e.g. flares \citep{reames_ssr_1999}. One important feature of impulsive SEP events is enhancement (factor of 3-10) of heavy ion abundance such as Ne/O and Fe/O (ratio of neon or iron to oxygen ion fluxes). Furthermore, it was also found that the abundances of ultra heavy ions with mass number up to 200 in impulsive events are also greatly enhanced over their solar abundances. The enhancement factor scales roughly as a power function of the change to mass ratio, i.e. $\propto(q/m)^p$, with a power index $p=-3.26$ \citep{mason_apj_2004}.
A comprehensive model of particle energization for impulsive SEP events must be able to reproduce this highly significant observation of heavy ion enhancement, as well as the well-known $\rm ^3 He$ enhancement (impulsive events are sometimes called $^3 \rm He$-rich events) \citep{mason_ssr_2007}. 

Among  theories proposed to explain particle heating and acceleration in impulsive SEP events, most of them rely on ubiquitous MHD turbulence. For example, \citet{miller_ssr_1998} showed that via resonant wave-particle interactions, ions (H, $^4 \rm He$, C, N, O, Ne, Mg, Si, Fe) can be stochastically accelerated by broadband Alfv\'en waves and fast mode waves. The enhancement of $^3 \rm He$ was produced separately by waves excited by an electron beam,  following the idea originally proposed by \citet{fisk_apj_1978}. 
\citet{liu_apj_2004,liu_apj_2006} built a model that could fit quantitatively the spectra of  $^3 \rm He$ and $^4 \rm He$ observed by ACE satellite, by including parallel Alfv\'en waves with a power-law spectrum, Coulomb losses, and diffusive escape of particles. At small scales, the Alfv\'en waves are assumed to evolve into proton cyclotron waves and helium cyclotron waves interacting with $^3 \rm He$ and $^4 \rm He$ differently, which leads to the enhancement of $^3 \rm He / ^4 \rm He$ ratio. In these models, the turbulence was assumed to be one dimensional, i.e., Alfv\'en waves were assumed to be propagating along the background magnetic field and fast modes were assumed to be isotropic.
However, recent development of MHD turbulence theory showed that turbulence in a magnetized plasma becomes strongly anisotropic as it evolves, with fluctuation energy mainly in the perpendicular direction \citep{sheba_jpp_1983,gs95}. The anisotropic turbulence has been confirmed by numerical simulations~\citep[e.g.,][]{cho_apj_2000,maron_apj_2001} and broadly observed in solar wind \citep[see][for a recent comprehensive review]{chen_jpp_2016}. This anisotropy may strongly affect stochastic ion heating and need to be modeled properly. 

Most of the studies of plasma turbulence assume plasma beta to be around unity, which is the typical value of solar wind near 1 AU. In this parameter regime, compressible fluctuations are deemed secondary~\citep{zank_pof_1993} and their effects on particle acceleration are negligible. When plasma beta is low ($<0.1$) and magnetic fluctuations remains high, compressible mode can play an important role in particle energization. For example, parametric decay instability can be triggered and cause the conversion of Alfv\'en waves into slow mode \citep{derby_apj_1978, golds_apj_1978}, even in a turbulent background \citep{shi_apj_2017}. Simulations have shown that subsequent damping of the slow mode can heat protons significantly \citep{fu_apj_2018}. Theoretical analysis by \citet{chand_prl_2005} showed that three-wave interactions can also transfer energy from low-frequency Alfv\'en waves to high-frequency fast waves, which can potentially explain the anisotropic heating of minor ions in the solar corona.

In this work, we focus on ion heating by turbulence developed in low-beta plasmas using 3D hybrid simulation. With 3D simulations turbulence can fully develop. Compared to MHD models, the hybrid model captures ion kinetic effects and self-consistent ion heating. Compared to fully kinetic particle-in-cell simulations, the hybrid model has the advantage of extending into larger spatial scale (in the inertial range). As we will see in following sections, turbulence with a power-law spectrum is capable of heating minor ions, producing temperature enhancement as a function of $q/m$. The heating mechanism is cyclotron resonance with nearly perpendicular  compressible waves.  We also compare the simulations to the stochastic heating model of $^3 \rm He$ based on 1-D turbulence \citep{liu_apj_2004,liu_apj_2006}. Our results may explain the observed dependence of the enhancement factor for heavy ions in impulsive SEP events, as reported by \citet{mason_apj_2004}. 

The paper is organized as follows. In Section \ref{sec:sim}, description of the hybrid code and the parameters of the numerical simulations are given. Results of the simulations are presented in Section \ref{sec:res}. Finally, summary of the main results, discussions of their applications and limitations of our model are given in Section \ref{sec:dis}.

\section{\label{sec:sim}Simulation Model}
A massively parallel 3D hybrid code -- H3D is used in this study \citep{karim_asp_2006, podes_jgr_2017, fu_apj_2018}. We treat ions as marker particles in the traditional particle-in-cell fashion, and electrons as a massless fluid. Focusing on low-frequency fluctuations and ion kinetics, we assume quasineutrality and ignore the displacement current. The electric field is solved using the so-called ``ion velocity extrapolation'' method and the magnetic field is advanced with the 4th order Runge-Kutta method \citep{winske_book_1993}. Triply periodic boundary conditions are applied for both particles and fields. The electron fluid is modelled by an adiabatic equation of state $T_e/n_e^{\Gamma -1}=\rm const$, where $\Gamma = 5/3$ is the adiabatic index. A small uniform resistivity $\eta  = 4\pi\times 10^{-6}$ is used to suppress short-wavelength (close to the grid size) noises and a binomial smoothing of moments (density and flow velocity) is also applied.  
Total energy is typically conserved with a relative error of a few percents in all cases presented here.

\begin{deluxetable}{cccccccc}
  \tablecaption{Key parameters for 3D hybrid simulations. $n_j$ is the
   relative density of species $j$ to the electron density, and the ion species is indicated in parentheses.\label{tab:para}}
  \tablecolumns{8}
  \tablehead{
  \colhead{Run} &
  \colhead{number of cells} &
  \colhead{$\beta_j$} &
  \colhead{$n_1$($\rm H^+$)} &
  \colhead{$n_{2}$($^4\rm He^{2+}$)} &
  \colhead{$n_{3}$($^3\rm He^{2+}$)} &
  \colhead{$n_4$($^{56}\rm Fe^{20+}$)} &
  \colhead{$n_5$($^{16}\rm O^{7+}$)}
  }
  \startdata
     1 & $240\times 240\times 240$& 0.05&0.90&0.09985&$3\times 10^{-5}$&$5\times 10^{-5}$&$7\times 10^{-5}$\\
     2 & $540\times 540\times 540$& 0.05&0.90&0.09985&$3\times 10^{-5}$&$5\times 10^{-5}$&$7\times 10^{-5}$\\
    3 & $240\times 240\times 240$& 0.005&0.90&0.09985&$3\times 10^{-5}$&$5\times 10^{-5}$&$7\times 10^{-5}$\\
     4 & $240\times 240\times 240$& 0.5&0.90&0.09985&$3\times 10^{-5}$&$5\times 10^{-5}$&$7\times 10^{-5}$\\
     5 & $240\times 240\times 240$& 0.005&0.9998&$5\times 10^{-5}$&$5\times 10^{-5}$&$5\times 10^{-5}$&$5\times 10^{-5}$\\
  \enddata
\end{deluxetable}

Key parameters for our 3D hybrid simulations are summarized in Table 1. The simulation domain is a cube of size $L^3$.
Aiming to understand ion heating in impulsive SEP events, which likely occur in the low-beta solar corona, we choose to study a proton-helium-electron plasma with $\beta$ on the order of 0.01-0.1 (for reference, $\beta=\beta_i+\beta_e \sim 0.01$ if $n=10^{10}$ /cc, $T_i=T_e= 100$ eV and $B_0= 100$ G). We model plasma turbulence developed by nonlinear interactions of low-frequency shear Alfv\'en waves possibly driven by magnetic reconnection in a solar flare. In the simulations, we initiate this process by loading 3 pairs of low-frequency long-wavelength counter-propagating Alfv\'en waves at $t=0$, each of which has an amplitude $a_0$ \citep{fu_apj_2018}:
\begin{eqnarray}
  \delta {\bf B}/B_0&=&\sum_{j,k}a_0\cos(jk_0 y+kk_0
  z+\phi_{j,k})\hat{\bf x}\nonumber\\
  &&+\sum_{l,n}a_0\cos(lk_0 x +n k_0 z+\phi_{l,n})\hat{\bf y}
\end{eqnarray}
\begin{eqnarray}
  \delta {\bf v}/v_A&=-&\sum_{j,k}a_0{\,\rm sgn}(k)\cos(jk_0 y+kk_0
  z+\phi_{j,k})\hat{\bf x}\nonumber\\
  &&-\sum_{l,n}a_0{\,\rm sgn}(n)\cos(lk_0 x+n k_0 z+\phi_{l,n})\hat{\bf y}
\end{eqnarray}
where  $(j,k)= (1,1), ( 2,1), ( 3, -2)$, $(l,n)= (1,-1),(-2,
-1),(-3, 2)$, $k_0=2\pi/L$ and the phase of each wave $\phi$ is random. The domain size is $L=240 d_i$ in most of the cases. We choose the amplitude of each wave ($a_0$) to be 0.1, resulting in magnetic fluctuation with root mean square $\delta B^{\text rms}/B_0\sim 0.24$ at t=0.
This type of simulation is often termed ``decaying turbulence''. We employ several ion species typically observed in SEP events, with protons and $^4\text{He}^{2+}$ as the major component dominating the dynamics and other components ($^3\text{He}^{2+}$, $^{16}\text{O}^{7+}$ and $^{56}\text{Fe}^{20+}$) as minor or trace components. Heavy ions, initially having the same temperature as protons, will be interacting with the electromagnetic fluctuations as the turbulence develops. It is worth noting that in this parameter regime, i.e. low beta and strong magnetic fluctuations, the turbulence Mach number $M\equiv\delta v/c_s$ is close to unity and the flows are compressible, in contrast to nearly incompressible turbulence typically observed near the Earth \citep{zank_pof_1993}.

\section{\label{sec:res}Results}
\begin{figure}
   \centering
   \includegraphics[width=0.6\textwidth]{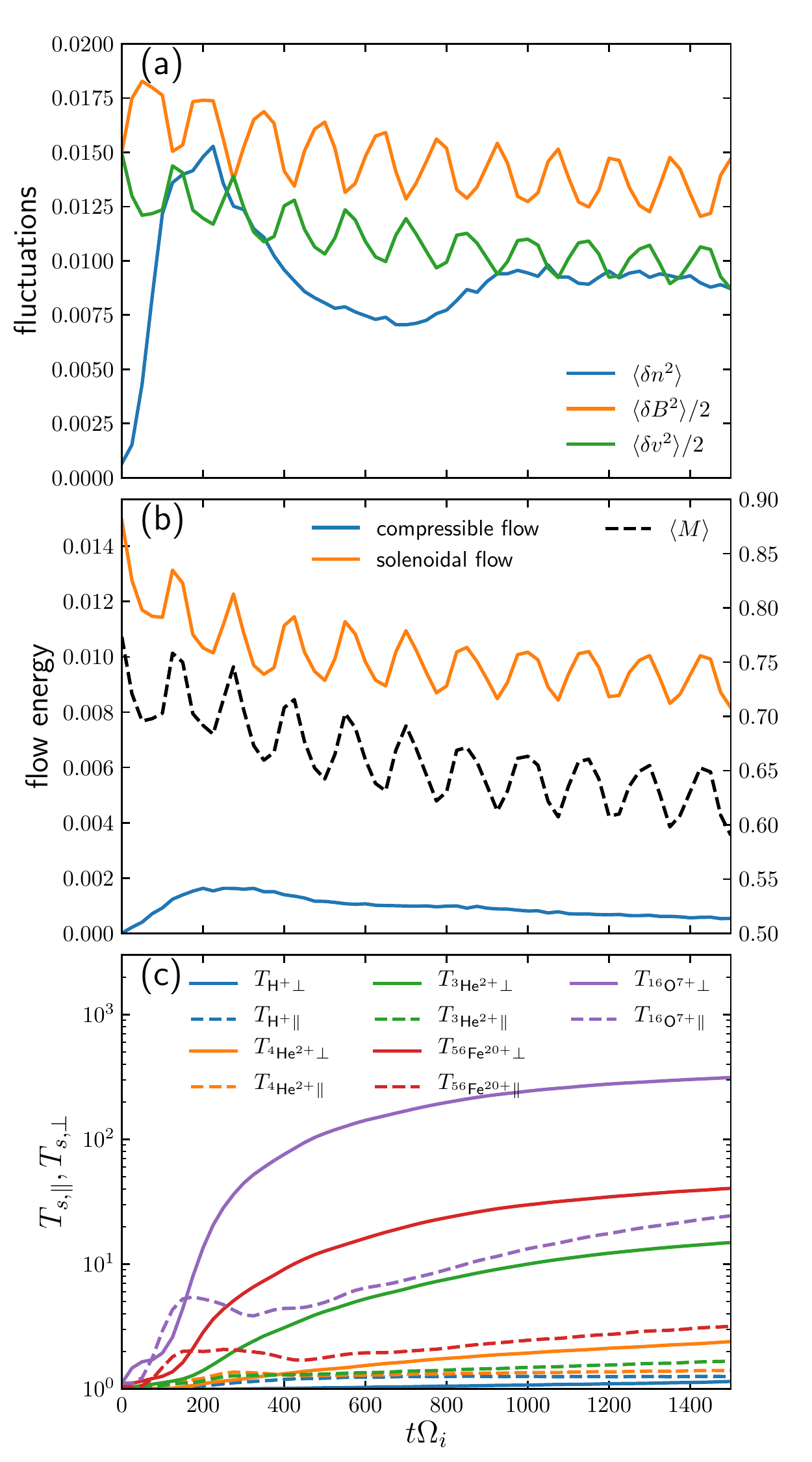}
   \caption{Evolution of (a) normalized density, flow velocity and magnetic field fluctuations, (b) energies of compressible flow and solenoidal flow (left axis), averaged turbulence Mach number $M\equiv \delta v/c_s$ (right axis),  and (c) perpendicular and parallel temperatures of different ion species in decaying turbulence Run 1.}
 \label{fig:history}
 \end{figure}

Figure \ref{fig:history} shows the time history of various physical quantities averaged over the simulation domain in Run 1. As a decaying turbulence simulation, the ion flow energy $(\delta v/v_A)^2/2$ and the fluctuating magnetic field energy $(\delta B/B_0)^2/2$ decrease gradually throughout the run, converting into plasma energy, as shown in Figure  \ref{fig:history}(a). {The oscillation of fluctuation energies (with a frequency $\omega\sim 2 k_0 v_A$) is due to nonlinear interaction of multiple waves that generates second harmonics of the fundamental mode.} At the end of the simulation ($t\Omega_i=1500$), about 20\% of the fluctuating magnetic energy  and 30\% of flow energy injected at $t=0$ have been converted, resulting in $\sim$ 25\% increase of average ion thermal energy. The density fluctuation experiences an exponential growth before $t\Omega_i=200$, due to parametric decay instability (PDI) of large-amplitude Alfv\'en waves in the low-$\beta$ environment \citep{fu_apj_2018}. The PDI converts a forward propagating Alfv\'en wave into a backward propagating Alfv\'en wave and a forward propagating ion acoustic wave. These ion acoustic waves have long wavelengths comparable to those of Alfv\'en waves ($\sim 100 d_i$).

The density fluctuation starts to decrease after $t\Omega_i=200$, due to Landau damping of ion acoustic waves causing parallel heating of major ions (dashed lines in Figure \ref{fig:history}b), similar to our previous study \citep{fu_apj_2018}. A second hump of $\delta n^2$ near $t\Omega_i=1000$ is due to the excitation of fast magnetosonic mode (discussed later). 

Throughout the simulation the density fluctuations stay high (8\% - 12\%), which is a signature of compressible turbulence. We can decompose the velocity field into a compressible component and a solenoidal (incompressbile) component, using Helmholtz decomposition
$$
{\bf v}=-\nabla\phi +\nabla\times{\bf A},
$$ where $\phi$ is a scalar field and ${\bf A}$ is a vector field.
As shown in Figure~\ref{fig:history}(b), the flow is purely incompressible at t=0 because Alfven waves are incompressible. Then compressible flow starts to grow and reaches its maximum around $t\Omega_i=200$, when the energy of compressible component is about 14\% of the total flow energy. It decays slowly afterwards and comprises 6\% of the total energy at the end of the simulation. Another indicator of the compressiblity is the turbulence Mach number $M$, which remains high (between 0.6 and 0.8) throughout the simulation. These compressible modes, as shown later, are responsible for strong heavy ion heating. 

\begin{figure}
   \centering
   \includegraphics[width=0.6\textwidth]{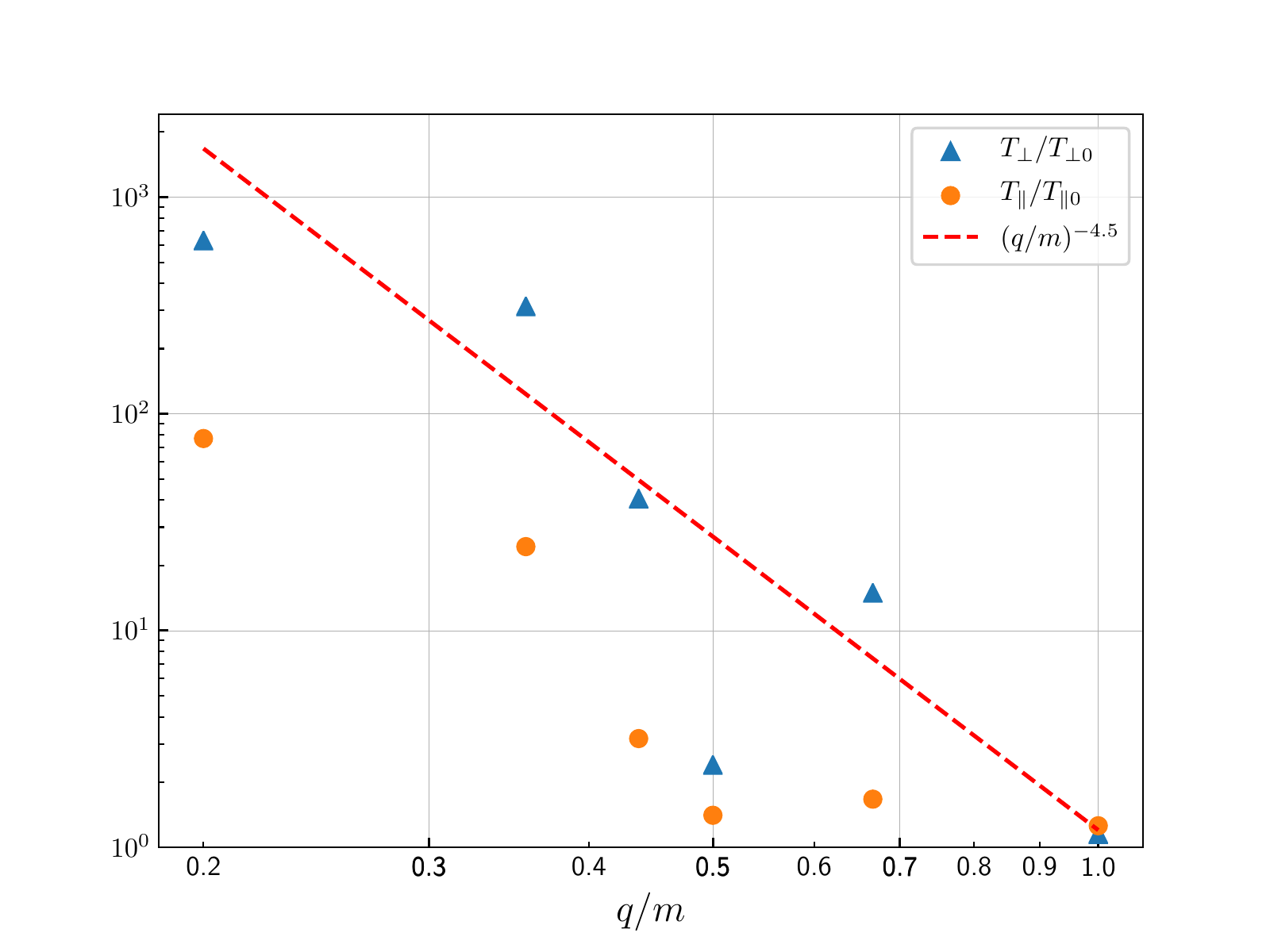}\\
   \caption{Temperature enhancement $T/T(t=0)$ as a function of the charge to mass ratio for all ion species at the end of the simulation of Run 1 (an extra heavy species with charge number 20 and mass number 100 is added here). The perpendicular temperature enhancement of ions scales roughly as $(q/m)^{-4.5}$, with $^4\rm He^{2+}$ ($q/m=0.5$) as an outlier. }
 \label{fig:scaling}
 \end{figure}

Figure \ref{fig:history}(c) shows the time history of perpendicular and parallel temperatures of all ion species in the simulation. Minor ions $^3\rm He^{2+}$ (species 3), $^{16}\rm O^{7+}$ (species 4), and $^{56}\rm Fe^{20+}$ (species 5) are significantly heated as the turbulence develops, with perpendicular temperature increase by a factor of 15, 40,  and 300, respectively. Since the density of minor ions are so low, they can keep extracting energy from the turbulence with little feedback to the system. In contrast, temperature enhancement for protons and $^4\rm He^{2+}$ is relatively small, although they absorb most of the energy. Note that the energy gain for all minor ion species is dominantly in the perpendicular direction, except for the very early stage ($t\Omega_i<200$) when PDI-generated ion acoustic waves are heating ions in the parallel direction \citep{fu_apj_2018}. At the end of the simulation ($t\Omega_i=1500$) the temperature anisotropy $T_\perp/T_\parallel$ of $^3\rm He^{2+}$, $^{16}\rm O^{7+}$, and $^{56}\rm Fe^{20+}$ reaches 8.9, 12.5, and 12.8, respectively. As shown in Figure \ref{fig:scaling}, the temperature enhancement (as measured by the ratio of temperature at $t\Omega_i=1500$ to that at $t=0$) of minor ions is inversely correlated with $q/m$, scaling roughly as $(q/m)^{-4.5}$.  

\begin{figure}
   \centering
   \includegraphics[width=0.9\textwidth]{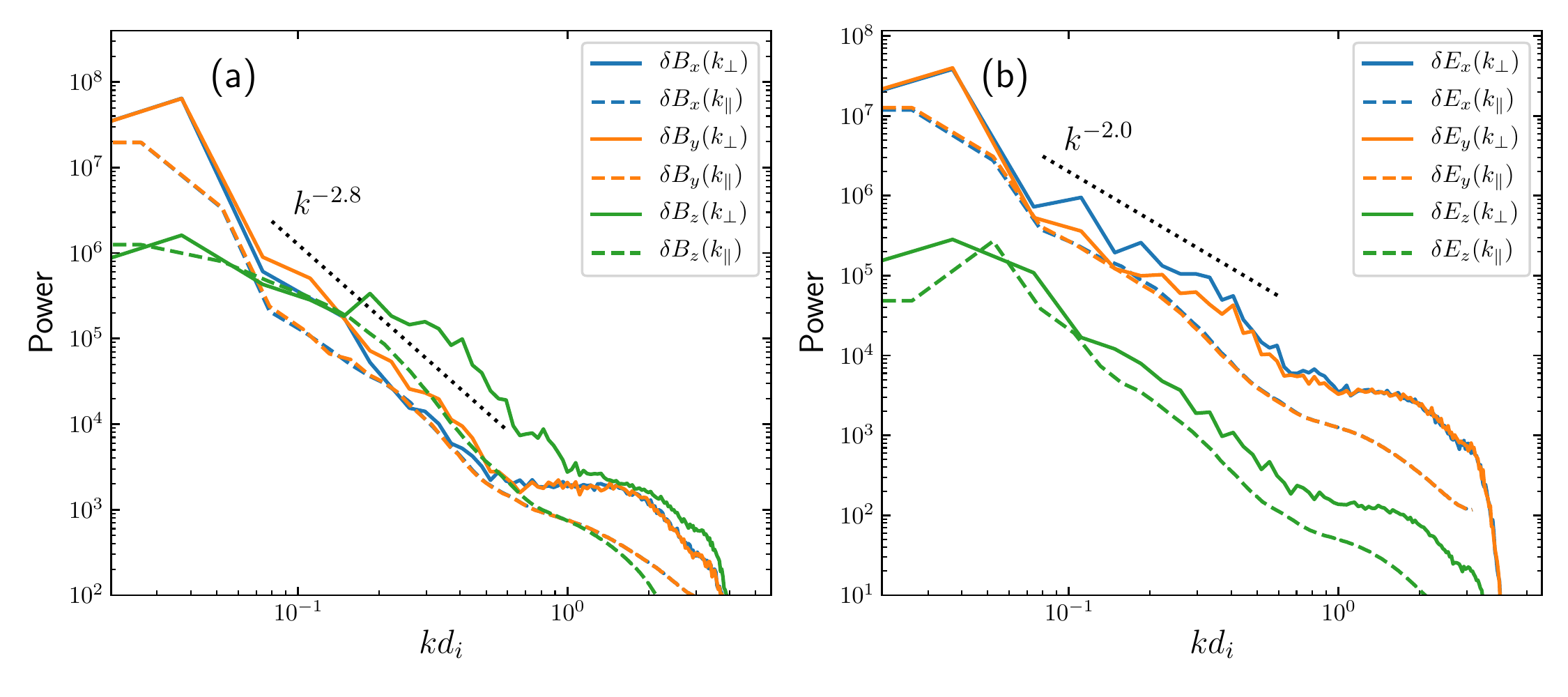}
   
   \caption{Power spectra of fluctuating (a) magnetic and (b) electric fields as a function of parallel wave number $k_\parallel$ (dashed lines) and perpendicular wave number $k_\perp$ (solid lines) for Run 1 at $t\Omega_i=500$. In the range $0.08<k d_i<0.6$, the energy density of perpendicular magnetic fluctuation and electric fluctuation scales as $k^{-2.8}$ and $k^{-2.0}$, respectively.  }
 \label{fig:spectra}
 \end{figure}

To study the nature of the generated turbulence, we calculate the power spectra of fluctuating magnetic and electric fields as a function of parallel and perpendicular wave numbers. Figure \ref{fig:spectra} shows the spectra at $t\Omega_i=500$ in Run 1, when the turbulence has fully developed and perpendicular ion heating is dominant (Figure \ref{fig:history}b). The turbulence is anisotropic, with more energy in the perpendicular waves modes (solid lines) than in the parallel modes (dashed lines). Transverse magnetic fields ($\delta B_x$ and $\delta B_y$) exhibit  a power law distribution in the inertial range ($0.03<k_\perp d_i<0.5$), which is typical for incompressible Alfv\'enic turbulence~\citep[e.g.][]{maron_apj_2001}. The spectrum obtained here has a power index close to -2.8. We have done some tests with stronger turbulence (e.g. increase $a_0$ to 0.2 or higher), a flatter spectrum with the index of -5/3 can be obtained, as predicted by the critical balance in strong turbulence theory \citep{gs95}. More interestingly,  the parallel magnetic fluctuations $\delta B_z$ is stronger than the perpendicular magnetic fluctuations in the the range  $0.2<k_\perp d_i<1.0$. This is a signature of compressible turbulence beyond the incompressible MHD framework. Perpendicular electric field fluctuations show a harder power law spectrum than that of the magnetic field, scaling as $k_\perp^{-2}$. But the parallel electric field is much weaker than the perpendicular electric field, indicating the fluctuations in the simulation are mainly electromagnetic.

\begin{figure}
   \centering
   \includegraphics[width=0.9\textwidth]{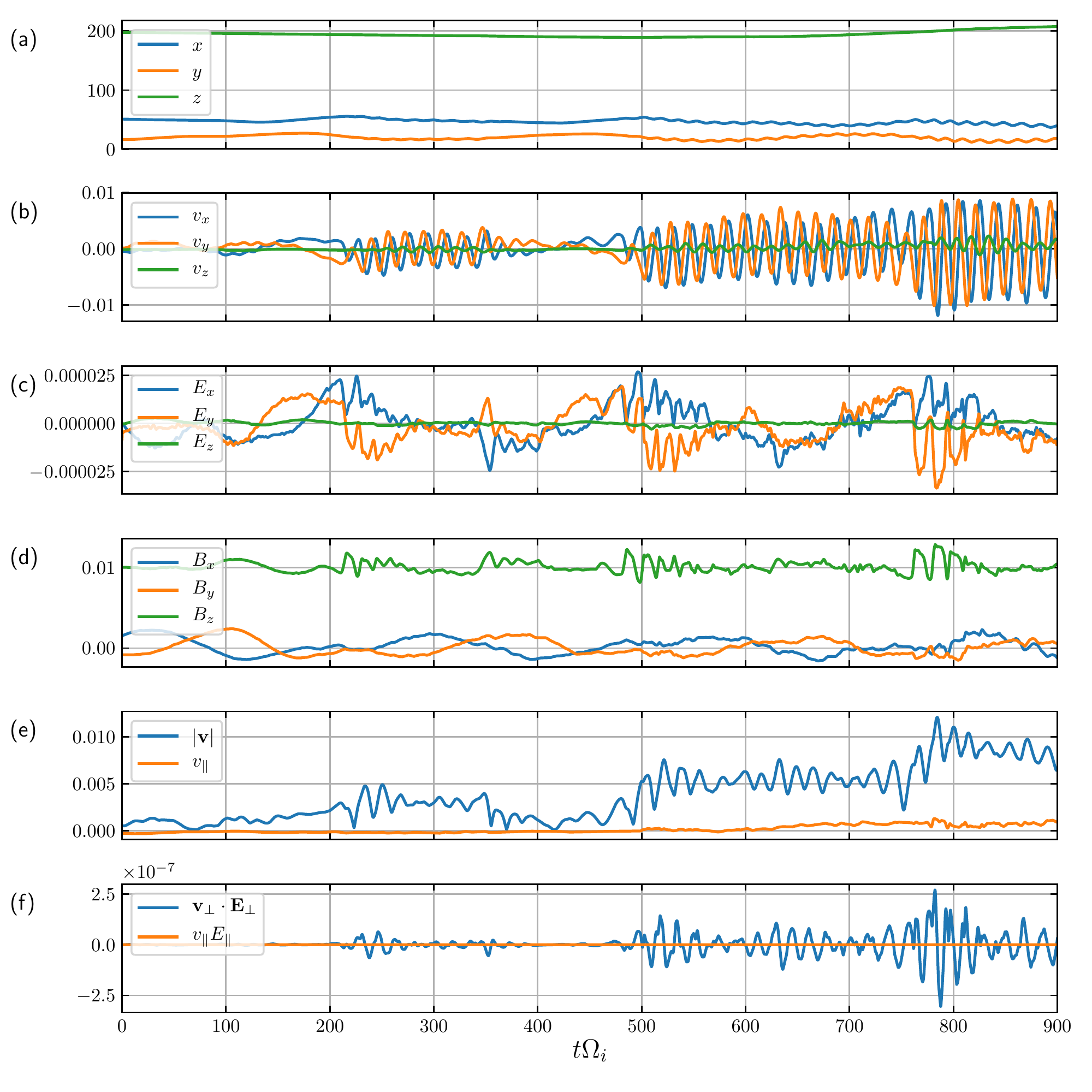}\\
   \caption{Time history of a tracked $\rm Fe^{20+}$ in Run 1: (a) position, (b) velocity, (c) electric fields seen by the particle, (d) magnetic fields seen by the particle, (e) total speed and parallel speed, and (f) work exerted by the parallel and perpendicular electric fields. The ion experience significant energy gain around $t\Omega_i=230, 500, 780$ and energy loss at $t\Omega_i=350$, by interacting with wave structures with enhanced $E_x$, $E_y$ and $B_z$ fluctuations.}
 \label{fig:particle}
 \end{figure}

To reveal the energization mechanism for ions, we randomly track 1600 particles for each species in the simulation, and then pick 200 particles with higher energies at the end of the simulation to analyze. Figure \ref{fig:particle} shows the time history of a tracked $\rm Fe^{20+}$ particle. In the first part of the simulation from $t=0$ to $t\Omega_i=900$, the ion experience three clear episodes of energy gain at around $t\Omega_i=230$, 500, and 780, respectively (Panel e). The energy gain is mostly in the perpendicular direction because the change of $v_\parallel$ is negligible, which is also confirmed in Panel f by the work exerted by parallel and perpendicular electric field. In each of these episodes, the particle encounters enhanced fluctuations featured by $E_x$, $E_y$ (Panel c) and $B_z$ (Panel d). Take the episode around $t\Omega_i=500$ for example, within a few gyroperiods the particle energy increases by a factor of 4, with essentially no change of parallel speed.  Electric and magnetic fields have a frequency close to its gyrofrequency ($\sim 0.36 \Omega_i$).  During this period of time, the ion moves in the region around $x/d_i=54, y/d_i=21, z/d_i=189$ and sees fluctuating electric fields $E_x$ and $E_y$ and magnetic field $B_z$ whose frequency is close to the ion gyrofrequency.

\begin{figure}
   \centering
   \includegraphics[width=0.9\textwidth]{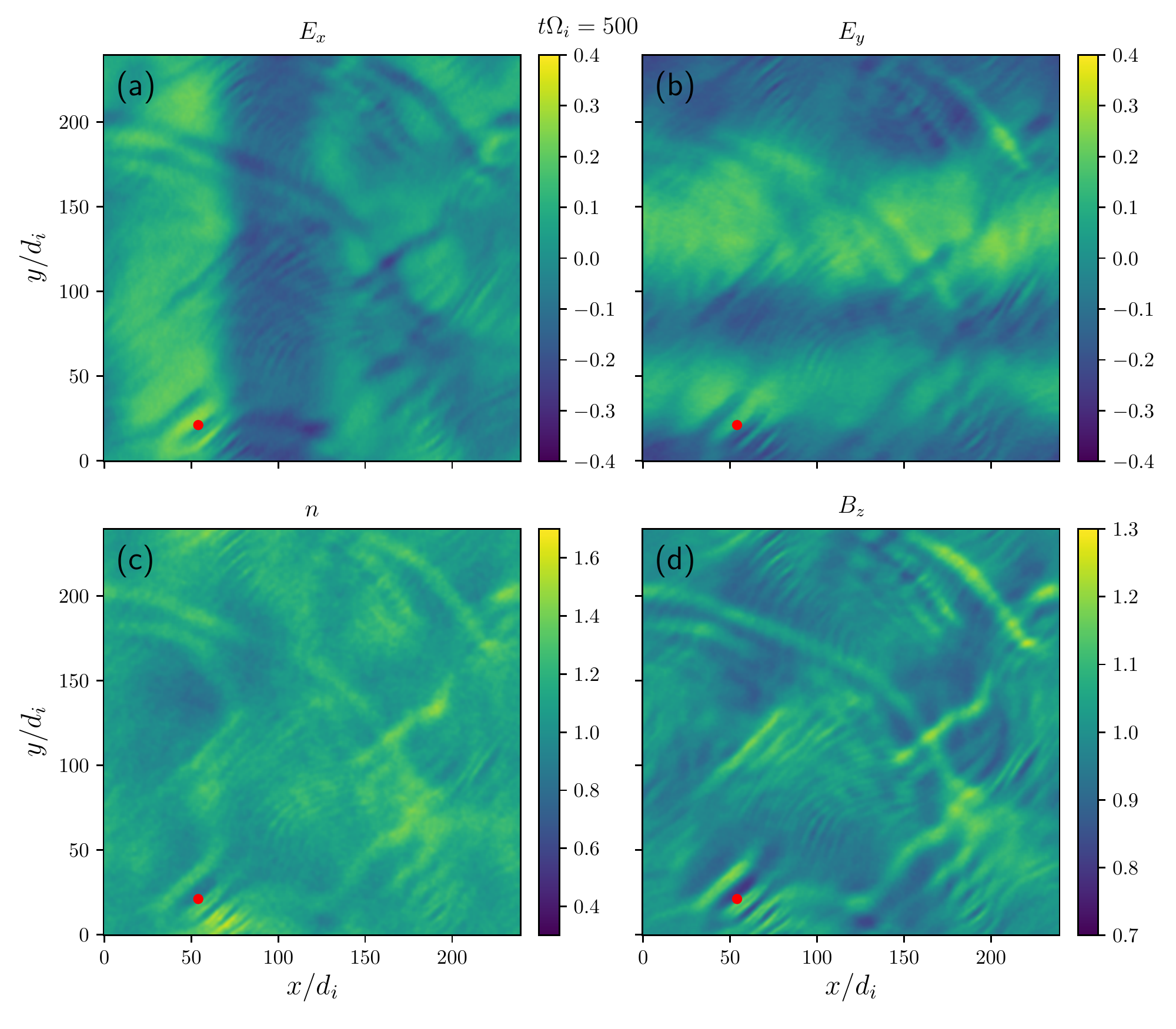}\\
   \caption{Contours of (a) electric field component $E_x$, (b) $E_y$, (c) number density $n$ and (d) magnetic field component $B_z$ in the plane of $z=189$ at $t\Omega_i=500$ in Run 1. The particle in Fig. \ref{fig:particle} is located at $x=54$ and $y=21$ (indicated by the red dot), interacting with a nearby wave structure.}
 \label{fig:fields}
 \end{figure}

\begin{figure}
   \centering
   \includegraphics[width=0.9\textwidth]{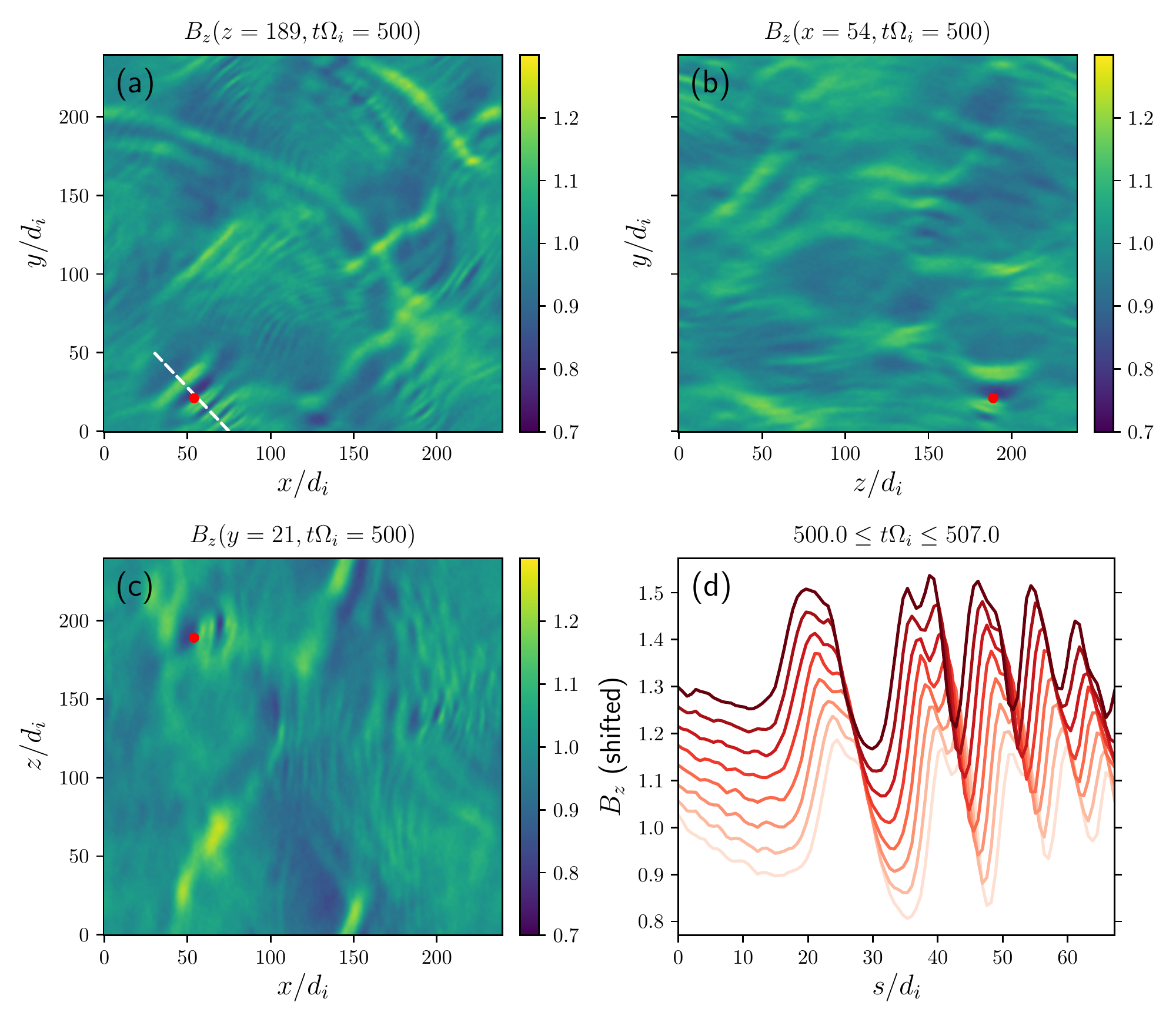}\\
   \caption{2D contours of $B_z$ in (a) $x-y$ plane, (b) $y-z$ plane, and (c) $x-z$ plane cut through the particle location $x=54, y=21, z=189$ (red dots). The wave structure has much stronger variation in the perpendicular direction (in $x-y$ plane) than in the parallel direction (along $z$). (d) The magnitude of $B_z$ along the wave normal direction (white dashed line in Panel a, from $x=35, y=50$ to $x=75, y=0$) from $t\Omega_i=500$ to $t\Omega_i=507$, with curves shifted up by 0.05 every $\Delta t = \Omega_i^{-1}$. The wave structure is estimated to propagate to upper left at a speed $\sim 0.8 v_A$.}
 \label{fig:bz}
 \end{figure}

These fluctuations are part of a localized wave structure that is shown in  
Figure \ref{fig:fields}. Contour of fields $E_x$, $E_y$, $B_z$ and number density $n$ in the $x-y$ plane at $z=189$ and $t\Omega=500$ are plotted, showing not only the large scale fluctuations (box size), but also many fine structures ($\sim 10 d_i$). The particle (indicated by the red dot) encounters the wave structure around $x=54, y=21$, which also has enhanced density fluctuations. 
To further examine its 3D structure, in Figure \ref{fig:bz} we plot 2D contours of $B_z$ in three planes cutting through the particle location (red dot). Clearly, the wave structure has a strong variation in the perpendicular direction (Fig. \ref{fig:bz}a), but a weak variation along the parallel direction (Fig. \ref{fig:bz}b and \ref{fig:bz}c), i.e. the wave number ${\bf k}\approx {\bf k}_\perp$. The structure is also localized, with finite extent in all three directions. Figure \ref{fig:bz}d shows the profiles of $B_z$ along the white dash line in Panel a (from $x=35, y=50$ to $x=75, y=0$), which is along the wave number direction, from $t\Omega_i=500$ to $t\Omega_i=507$. Profiles after $t\Omega_i=500$ have been shifted up by 0.05 every $\Delta t\Omega_i=1$. From this time stack plot, we estimate the wave number $0.48<k_\perp d_i<0.90$ and the phase speed $v_p\approx 0.8 v_A$. 
The properties of the structure are consistent with those of highly oblique fast magnetosonic (MS) mode having a dispersion relation $\omega=k_\perp v_A$. This is further supported by the fact that $\delta B_z $ ($\sim \delta B_\parallel$) fluctuation is correlated well with the density fluctuation $\delta n$ (Fig. \ref{fig:fields}c), a characteristic of the compressible MS wave.

The energization process can be understood as resonant wave-particle interaction. The cyclotron resonance condition for wave-particle interaction
$\omega-k_\parallel v_\parallel=n\Omega_i$ (where $n=0, \pm 1, \pm 2, \dots$) is reduced to 
\begin{equation}
    \omega=n\Omega_i
\end{equation}
for perpendicular waves.
To illustrate the interaction of ions with nearly perpendicular MS waves, we consider a monochromatic plane MS wave in a uniform plasma, whose dispersion relation is given by 
$ \omega=k_\perp v_A$. Assuming $k_\perp=k_x$ and ${\bf B}_0=B_0\hat{\bf e}_z$, the fluctuating electric and magnetic fields are given by 
\begin{equation}
    \delta B_z/B_0=\epsilon \cos (k_x x-\omega t),
\end{equation}
\begin{equation}
cE_y/B_0v_A=-\epsilon \cos(k_x x-\omega t).
\end{equation}

We then follow the motion of an ion (gyrofrequency $\Omega_i$) in such a wave by solving Newton's equation with the Lorentz force. 

\begin{figure}
   \centering
   \includegraphics[width=0.9\textwidth]{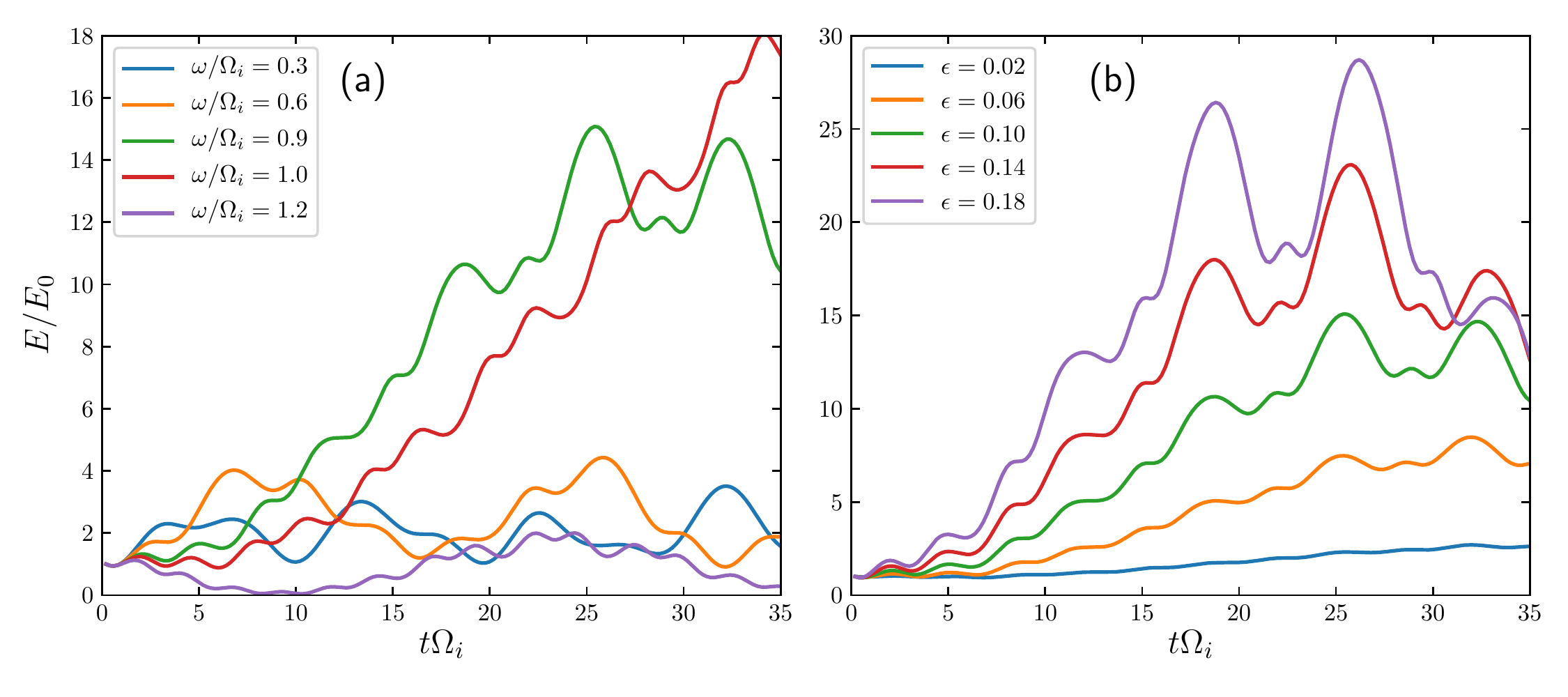}
    
   \caption{Energy evolution of an ion (cyclotron frequency $\Omega_i$) interacting with an ideal monochromatic magnetosonic wave (frequency $\omega$, amplitude $\epsilon$) propagating perpendicular to the background magnetic field: (a) with waves of different frequencies but fixed amplitude $\epsilon=0.1$, (b) with waves of different amplitude but fixed frequency $\omega/\Omega_i=0.9$.}
 \label{fig:mswave}
 \end{figure}

Figure \ref{fig:mswave} shows the evolution of ion energy in the presence of waves with different frequencies and amplitudes. The initial particle beta is $0.01$. In Fig. \ref{fig:mswave}(a) the wave amplitude is fixed at $\epsilon=0.1$. When the wave frequency is close to the ion gyrofrequency ($0.3< \omega/\Omega_i< 1.2$), the particle can quickly gain energy within several gyroperiods.  The closer the wave frequency is to the ion gyrofrequency, the longer time the ion can be in phase with the wave and more energy is transferred from the wave to the particle. For example, the particle's energy increases by more than a factor of 10 when interacting with a wave with $\omega/\Omega_i=0.9$ within about 25$\Omega_i^{-1}$ or 4 gyroperiods.
Since we assume a plane wave in an infinite space, the particle will lose its energy eventually and return to its initial state due to the periodicity, except when $\omega$ is exactly at $\Omega_i$. The period of returning to the initial state is inversely proportional to the frequency difference $\Delta\omega=\omega-\Omega_i$. It also means that the resonant  wave-particle interaction is a random process -- ions can gain energy or loss energy depending on the phase difference between the particle gyro-motion and the wave. In fact, it is directly observed in our simulation, as in Figure \ref{fig:particle}, that the ion loses its energy by interacting with the MS wave structure near $t\Omega_i=370$.  
In Fig. \ref{fig:mswave}(b) the wave frequency is fixed at $\omega/\Omega_i=0.9$ and the maximum energy gain is proportional to the wave amplitude. This dependence explains the different temperature enhancements for different trace ion species shown in Fig. \ref{fig:history}b.
Each ion species $j$ has a different charge to mass ratio $q_j/m_j$ such that its gyrofrquency normalized to proton gyrofrequency $\Omega_j/\Omega_i=q_jm_i/q_im_j$ is 0.36 for $^{56}\rm Fe^{20+}$, 0.44 for $^{16}\rm O^{7+}$, and 0.67 for $^3\rm He^{2+}$. The spectra of the turbulent fluctuations in Fig. \ref{fig:spectra} show that energy density is lower in higher frequencies (the horizontal axis $kd_i$ can be translated into frequency by $\omega\sim kv_A$), and therefore $^{56}\rm Fe^{20+}$ ions are heated most and $^3\rm He^{2+}$ ion are heated least. This does not apply to $^4\rm He^{2+}$ because it is a major component of the plasma and there is insufficient energy to heat them all to very high temperature. In a separate simulation where $^4\rm He^{2+}$ is of trace amount (Run 5),  $^4\rm He^{2+}$ is heated strongly like other species (not shown).
 
So far we have been focusing on the energization process of one particular ion at a particular time. It should be pointed out that this process is typical for most of the 200 particles (including all three trace ion species) we have analyzed, in which ions exchange energy with compressible MS wave structures multiple times. Overall, it leads to significant heating of trace ion species. 

To establish the reliability of our simulation results, we have done a few convergence tests. A higher resolution run with more than double grid points in each direction (Run 2 in Table \ref{tab:para}) yields very similar results. Another run with 10 times more particles than Run 1 (with other parameters unchanged) confirms that the numerical noise common in particle-in-cell simulations does not affect the physics we are studying.

Finally, we study the dependence of ion heating on the plasma beta. The initial ion beta is reduced by a factor of 10 in Run 3 compared to Run 1. With the injected wave energy fixed, this effectively changes the turbulent Mach number $M\sim 2.2$. The resulting density fluctuation  $\delta n^2$ at $t\Omega_i=1500$ reaches 0.1 which is about 10 times larger than that in Run 1 and the heating of heavy ions (measured by the ratio of the final temperature to the initial temperature) are also about 10 times stronger. In Run 4, the plasma beta is increased by a factor of 10 and $M$ is reduced to 0.2. The density fluctuation is reduced by a factor of $\sim 6$ and the heating for heavy ions are reduced by a factor of $\sim 12$.

\section{\label{sec:dis} Summary and Discussion}
In this paper, we report results from 3D hybrid simulations on ion heating in a highly turbulent low-beta plasma where the turbulent Mach number is close to unity. It is shown that injection of large-scale Alfv\'en waves develops into compressible and anisotropic turbulence, which efficiently heats heavy ion species through cyclotron resonance. Temperature enhancement of minor ions is inversely proportional to the charge to mass ratio because heavier ions have access to lower frequency fluctuations that have higher energy density. This result can explain heavy ion enhancement typically observed in impulsive SEP events. Further analysis confirms that ions are energized by interacting with nearly perpendicular magnetosonic waves on the scale of several proton inertial lengths. Since the compressible waves play a key role in ion energization, this process is more robust in the regime of high turbulent Mach number, such as the close-to-the-Sun region. 

This heating process is quite different from cyclotron heating of ions by parallel Alfv\'enic fluctuations or isotropic magnetosonic turbulence \citep{ miller_ssr_1998, liu_apj_2004,liu_apj_2006}. The turbulence is highly anisotropic, with majority of the energy in the perpendicular direction.
While $^3\rm He^{2+}$ particles included in our simulations are energized more than $^4\rm He^{2+}$, the amount of heating is not enough to explain  the preferential heating of $^3\rm He^{2+}$ observed in impulsive SEP events~\citep{mason_ssr_2007}. This may suggest that the cyclotron heating model of $^3\rm He^{2+}$ due to 1D wave cascade is oversimplified \citep{liu_apj_2004,liu_apj_2006}. To explain the preferential heating observed $^3\rm He^{2+}$ in SEP events, additional physics or energy source is required, such as an electron beam or temperature anisotropy that can drive additional ion cyclotron waves that heats Helium particles \citep[e.g.,][]{temer_apj_1992,miller_ssr_1998}.  

In our simulations, we inject Alfven waves with $\delta B^{\text rms}/B_0=0.24$ at scale $0.03<kd_i<0.1$. The amplitude may be larger than what we expect at this scale near the Sun where impulsive SEPs are produced. 

Large amplitude energy injection is needed to overcome numerical noise in the code. Another limitation is that the scale separation may be insufficient to model the turbulence in real plasma. For example, the fast modes generated in our simulation are close to the injection scale, and are also not too far from the kinetic scale. To confirm that they can cascade down to the kinetic scale (which is several orders of magnitude smaller than the injection scale in reality) where heavy ions are heated needs much larger scale separation, which is very challenging for current simulation codes. Lastly, generation and evolution of fast mode turbulence and its interactions with other MHD modes in different plasma environment (e.g. low and high beta) is a very interesting topic \citep[e.g.][]{cho_mnras_2003, chand_prl_2005, svidz_pop_2009}. But it is beyond the scope of the current study and is left for future investigation.

\acknowledgments 
The Los Alamos portion of
this research was performed under the auspices of the U.S. Department of
Energy. We are grateful for support from the LANL/LDRD program and DOE/OFES.
This research used resources provided by the Los Alamos National Laboratory Institutional Computing Program, which is supported by the U.S. Department of Energy National Nuclear Security Administration under Contract No. 89233218CNA000001. XF thanks Vadim Roytershteyn and Patrick Kilian for help with the hybrid code.


\begin{thebibliography}{}
\expandafter\ifx\csname natexlab\endcsname\relax\def\natexlab#1{#1}\fi
\providecommand{\url}[1]{\href{#1}{#1}}
\providecommand{\dodoi}[1]{doi:~\href{http://doi.org/#1}{\nolinkurl{#1}}}
\providecommand{\doeprint}[1]{\href{http://ascl.net/#1}{\nolinkurl{http://ascl.net/#1}}}
\providecommand{\doarXiv}[1]{\href{https://arxiv.org/abs/#1}{\nolinkurl{https://arxiv.org/abs/#1}}}

\bibitem[{Chandran(2005)}]{chand_prl_2005}
Chandran, B. D.~G. 2005, Physical Review Letters, 95, 265004,
  \dodoi{10.1103/PhysRevLett.95.265004}

\bibitem[{Chen(2016)}]{chen_jpp_2016}
Chen, C.~H. 2016, Journal of Plasma Physics, 82,
  \dodoi{10.1017/S0022377816001124}

\bibitem[{Cho \& Lazarian(2003)}]{cho_mnras_2003}
Cho, J., \& Lazarian, A. 2003, Monthly Notices of the Royal Astronomical
  Society, 345, 325, \dodoi{10.1046/j.1365-8711.2003.06941.x}

\bibitem[{Cho \& Vishniac(2000)}]{cho_apj_2000}
Cho, J., \& Vishniac, E.~T. 2000, The Astrophysical Journal, 539, 273,
  \dodoi{10.1086/309213}

\bibitem[{{Derby, Jr.}(1978)}]{derby_apj_1978}
{Derby, Jr.}, N.~F. 1978, The Astrophysical Journal, 224, 1013,
  \dodoi{10.1086/156451}

\bibitem[{Fisk(1978)}]{fisk_apj_1978}
Fisk, L.~A. 1978, The Astrophysical Journal, 224, 1048, \dodoi{10.1086/156456}

\bibitem[{Fu {et~al.}(2018)Fu, Li, Guo, Li, \& Roytershteyn}]{fu_apj_2018}
Fu, X., Li, H., Guo, F., Li, X., \& Roytershteyn, V. 2018, The Astrophysical
  Journal, 855, 139, \dodoi{10.3847/1538-4357/aaacd6}

\bibitem[{Goldreich \& Sridhar(1995)}]{gs95}
Goldreich, P., \& Sridhar, S. 1995, The Astronomical Journal, 438, 763,
  \dodoi{10.1086/174600}

\bibitem[{Goldstein(1978)}]{golds_apj_1978}
Goldstein, M.~L. 1978, The Astrophysical Journal, 219, 700,
  \dodoi{10.1086/155829}

\bibitem[{Karimabadi {et~al.}(2006)Karimabadi, Vu, Krauss-Varban, \&
  Omelchenko}]{karim_asp_2006}
Karimabadi, H., Vu, H.~X., Krauss-Varban, D., \& Omelchenko, Y. 2006, in
  Numerical Modeling of Space Plasma Flows, ASP Conference Series, ed. N.~V.
  Pogorelov \& G.~P. Zank, Vol. 359, 257--263.
\newblock \url{http://adsabs.harvard.edu/full/2006ASPC..359..257K}

\bibitem[{Liu {et~al.}(2004)Liu, Petrosian, \& Mason}]{liu_apj_2004}
Liu, S., Petrosian, V., \& Mason, G.~M. 2004, The Astrophysical Journal, 613,
  L81, \dodoi{10.1086/425070}

\bibitem[{Liu {et~al.}(2006)Liu, Petrosian, \& Mason}]{liu_apj_2006}
---. 2006, The Astrophysical Journal, 636, 462, \dodoi{10.1086/497883}

\bibitem[{Maron \& Goldreich(2001)}]{maron_apj_2001}
Maron, J., \& Goldreich, P. 2001, The Astrophysical Journal, 554, 1175,
  \dodoi{10.1086/321413}

\bibitem[{Mason(2007)}]{mason_ssr_2007}
Mason, G.~M. 2007, Space Science Reviews, 130, 231

\bibitem[{Mason {et~al.}(2004)Mason, Mazur, Dwyer, Jokipii, Gold, \&
  Krimigis}]{mason_apj_2004}
Mason, G.~M., Mazur, J.~E., Dwyer, J.~R., {et~al.} 2004, The Astrophysical
  Journal, 606, 555, \dodoi{10.1086/382864}

\bibitem[{Miller(1998)}]{miller_ssr_1998}
Miller, J.~A. 1998, Space Science Reviews, 86, 79,
  \dodoi{10.1023/A:1005066209536}

\bibitem[{Podesta \& Roytershteyn(2017)}]{podes_jgr_2017}
Podesta, J.~J., \& Roytershteyn, V. 2017, Journal of Geophysical Research:
  Space Physics, 122, 6991, \dodoi{10.1002/2017JA024074}

\bibitem[{Reames(1999)}]{reames_ssr_1999}
Reames, D.~V. 1999, Space Science Reviews, 90, 413

\bibitem[{Reames(2017)}]{reames_sp_2017}
---. 2017, Solar Physics, 292, 1, \dodoi{10.1007/s11207-017-1173-5}

\bibitem[{Shebalin {et~al.}(1983)Shebalin, Matthaeus, \&
  Montgomery}]{sheba_jpp_1983}
Shebalin, J.~V., Matthaeus, W.~H., \& Montgomery, D. 1983, Journal of Plasma
  Physics, 29, 525, \dodoi{10.1017/S0022377896005260}

\bibitem[{Shi {et~al.}(2017)Shi, Li, Xiao, \& Wang}]{shi_apj_2017}
Shi, M., Li, H., Xiao, C., \& Wang, X. 2017, The Astrophysical Journal, 842,
  63, \dodoi{10.3847/1538-4357/aa71b6}

\bibitem[{Svidzinski {et~al.}(2009)Svidzinski, Li, Rose, Albright, \&
  Bowers}]{svidz_pop_2009}
Svidzinski, V.~A., Li, H., Rose, H.~A., Albright, B.~J., \& Bowers, K.~J. 2009,
  Physics of Plasmas, 16, 1, \dodoi{10.1063/1.3274559}

\bibitem[{Temerin \& Roth(1992)}]{temer_apj_1992}
Temerin, M., \& Roth, I. 1992, The Astrophysical Journal, 391, L105

\bibitem[{Winske \& Omidi(1993)}]{winske_book_1993}
Winske, D., \& Omidi, N. 1993, in Computer Space Plasma Physics: Simulations
  and Software., ed. H.~Matsumoto \& Y.~Omura (Terra, Tokyo)

\bibitem[{Zank \& Matthaeus(1993)}]{zank_pof_1993}
Zank, G.~P., \& Matthaeus, W.~H. 1993, Physics of Fluids A: Fluid Dynamics, 5,
  257, \dodoi{10.1063/1.858780}

\end{thebibliography}

\end{document}